\documentclass[10pt,conference, letterpaper]{IEEEtran}
\IEEEoverridecommandlockouts
\usepackage{cite}
\usepackage{amsmath,amssymb,amsfonts}
\usepackage{algorithmic}
\usepackage{graphicx}
\usepackage{textcomp}
\usepackage{xcolor}
\def\BibTeX{{\rm B\kern-.05em{\sc i\kern-.025em b}\kern-.08em
    T\kern-.1667em\lower.7ex\hbox{E}\kern-.125emX}}

\usepackage{bbding}
\usepackage{pifont}
\usepackage{wasysym}
\usepackage{amssymb}

\usepackage[left=0.68in,top=0.78in,right=0.68in,bottom=0.78in]{geometry}
\newcommand{\etal}{\textit{et al.}}
\newcommand{\ie}{\textit{i.e.,~}}

\makeatletter
\let\old@ps@headings\ps@headings
\let\old@ps@IEEEtitlepagestyle\ps@IEEEtitlepagestyle
\def\confheader#1{
	\def\ps@IEEEtitlepagestyle{
		\old@ps@IEEEtitlepagestyle
		\def\@oddhead{\normalfont\scriptsize\centering#1}%
	}%
}

\makeatother

\confheader{%
	This paper has been accepted for publication
	in the IEEE Global Communications Conference (GLOBECOM'19), 09 December-13 December 2019, Waikoloa, HI, USA
}

\begin{document}

\title{LiSA: A Lightweight and Secure Authentication Mechanism for Smart Metering Infrastructure\\
}

\author{
	\IEEEauthorblockN{Sahil Garg\IEEEauthorrefmark{1}, Member, IEEE, Kuljeet Kaur\IEEEauthorrefmark{1}, Member, IEEE, Georges Kaddoum\IEEEauthorrefmark{1}, Member, IEEE, \\Fran\c{c}ois Gagnon\IEEEauthorrefmark{1}, Senior Member, IEEE, Syed Hassan Ahmed\IEEEauthorrefmark{2}, Senior Member, IEEE, \\ Dushantha Nalin K. Jayakody\IEEEauthorrefmark{3}, Senior Member, IEEE}
	\IEEEauthorblockA{\IEEEauthorrefmark{1}Electrical Engineering Department, \'Ecole de technologie sup\'erieure, Montr\'eal, QC H3C 1K3, Canada.\\
		\IEEEauthorrefmark{2}Department of Computer Science, Georgia Southern University, Statesboro, GA 30460, USA.\\
		\IEEEauthorrefmark{3}School of Computer Science and Robotics, National Research Tomsk Polytechnic University, Russia.\\}
	(e-mail: sahil.garg@ieee.org, kuljeet.kaur@ieee.org, georges.kaddoum@etsmtl.ca, francois.gagnon@etsmtl.ca, \\ sh.ahmed@ieee.org, nalin.jayakody@ieee.org)
}
\maketitle

\begin{abstract}	
Smart metering infrastructure (SMI) is the core component of the smart grid (SG) which enables two-way communication between consumers and utility companies to control, monitor, and manage the energy consumption data. Despite their salient features, SMIs equipped with information and communication technology are associated with new threats due to their dependency on public communication networks. Therefore, the security of SMI communications raises the need for robust authentication and key agreement primitives that can satisfy the security requirements of the SG. Thus, in order to realize the aforementioned issues, this paper introduces a lightweight and secure authentication protocol, \textit{``LiSA"}, primarily to secure SMIs in SG setups. The protocol employs Elliptic Curve Cryptography at its core to provide various security features such as mutual authentication, anonymity, replay protection, session key security, and resistance against various attacks. Precisely, LiSA exploits the hardness of the Elliptic Curve Qu Vanstone (EVQV) certificate mechanism along with Elliptic Curve Diffie Hellman Problem (ECDHP) and Elliptic Curve Discrete Logarithm Problem (ECDLP). Additionally, LiSA is designed to provide the highest level of security relative to the existing schemes with least computational and communicational overheads. For instance, LiSA incurred barely 11.826 ms and 0.992 ms for executing different passes across the smart meter and the service providers. Further, it required a total of 544 bits for message transmission during each session.
\end{abstract}


\section{Introduction}
{\color{black}
Smart Grid (SG) is the next-generation power system that greatly enhances the reliability, efficiency and sustainability of the legacy power systems with renewable energy sources, distributed intelligence, and improved demand response features \cite{kaur2018game, aujla2019drops}. As opposed to the traditional grid, SG allows a bi-directional flow of energy and information in order to enable new functionalities among the consumers and utilities. The smart metering infrastructure (SMI) is one of the crucial application domains of SGs that helps in evaluating the status of a power grid along with the management of the distributed resources \cite{8642293, 7959167}. In SGs, the smart meters (SMs) connect power consumers to the utility company in order to exchange, manage and control the energy consumption. They rely on advanced information and communication technologies to support the intelligent power supply and enhance the efficiency of legacy power systems\cite{kaursystem, kaur2019osco}. However, due to the lack of inherent and effective security mechanisms, interactions between legal entities are susceptible to cyber-attacks \cite{chaudhary2018sdn, 7093120}. Nevertheless, the complex nature of the SG and its diverse security requirements pose challenges to its widespread adoption. Thus, it is necessary to provide secure and reliable authentication systems for SMs that not only maintain trust between the legitimate entities but also satisfy other security services like mutual authentication, integrity, and anonymity \cite{2019arXiv190401171G, 8676005}. 

In order to address the privacy issues of SMs, several authentication schemes have recently been proposed. For example, Diao \textit{et al.} \cite{diao2015privacy} introduced a privacy preserving smart metering protocol that employs Camenisch–Lysyanskaya signature scheme for providing a secure and reliable authentication system. This scheme constructs linkable anonymous credential in order to provide message authentication and traceability of faulty SMs. In a similar context, Yu \textit{et al.} \cite{7138617} designed an information centric networking (ICN) based approach; wherein a novel key management scheme was employed in order to ensure the security in SMI. Similarly, Abbasinezhad-Mood and Nikooghadam \cite{7931655} devised a lightweight communication scheme for SMs which employs the hash and exclusive-OR operations in order to provide confidentiality, real-time authentication along with the ability to cope with one-minute or even less time intervals of data transmission. Likewise, the authors in \cite{seferian2018identity} employed an identity-based non-interactive key distribution scheme to develop a secure and scalable key distribution framework for neighboring SMs. In order to provide compression and authentication of smart meter readings in SMIs, Lee \textit{et al.} \cite{8662558} introduced a unified approach based on the notion of compressive sensing in a multicarrier system. Here, the residual error of a received signal was used to determine whether the signal is legitimate or not. 

In 2018, Mustapa \textit{et al.} \cite{mustapa2018hardware} proposed an authentication scheme for SMI; wherein ring oscillator physically unclonable functions (ROPUF) were employed for deriving and storing the cryptographic keys. In \cite{7676395}, Mohammadali \textit{et al.} utilized Elliptic Curve Cryptography (ECC) to propose an identity-based key establishment protocol for SMIs. Likewise, Kumar \textit{et al.} \cite{8413131} presented a lightweight authentication and key agreement scheme in order to maintain the privacy in SMIs. In an another work, Abbasinezhad-Mood and Nikooghadam \cite{8294238} developed an anonymous ECC-based self-certified key distribution scheme for SGs in order to solve the issues of public key infrastructure (PKI) maintenance and the key escrow problem. Similarly, Braeken \textit{et al.} \cite{braeken2018efficient} also presented a key agreement model for smart metering communications which aims to provide session key security under the widely accepted Canetti-Krawczyk security model. The authors asserted that their scheme provides identity-based mutual authentication, credential privacy, and session key security, along with the resistance against the well-known attacks such as replay, man-in-the-middle, and impersonation. Although several authentication and key exchange protocols exist in the literature to provide security in SGs, most of them not only fail to provide SM anonymity but also require high communicational and computational costs. Thus, in this paper a \textbf{L}ightweight and \textbf{S}ecure \textbf{A}uthentication (\textit{LiSA}) protocol is designed for securing SMIs.


\subsection{Contributions}
The key contributions of this paper are illustrated below:
\begin{itemize}
	\item We propose a provably secure  mutual authentication and key agreement protocol named \textit{LiSA}, specifically for SMIs employed in SG environments. The designed protocol leverages the hardness of the Elliptic Curve Qu Vanstone (EVQV) certificate mechanism, Elliptic Curve Diffie Hellman Problem (ECDHP), and Elliptic Curve Discrete Logarithm Problem (ECDLP) in its design to prevent various security attacks.
	\item Additionally, we provide a detailed informal security evaluation of the proposed protocol. The evaluation reveals that LiSA supports mutual authentication, anonymity, replay protection, session key security, and resists various other attacks.
	\item The proposed protocol is a perfect mix of security features and lightweight  cryptographic functions that make it apt for deployment in SG ecosystems. Thus, in order to validate it efficacy it has been extensively evaluated against the current state-of-the-art in terms of security features, computational complexity, and communicational complexity.
\end{itemize}
}

\subsection{Organization}
{\color{black}The present manuscript is structured according to the following sequence. The proposed system model is sketched in Section~\ref{sec:SystemModel} and the designed protocol is presented in Section~\ref{sec:ProposedProtocol}. Following this, detailed security analysis and performance assessment of the proposed scheme are demonstrated in Sections~\ref{sec:SecurityAnalysis} and \ref{sec:PerformanceEvaluation}, respectively. Lastly, the conclusions are drawn in Section~\ref{sec:Conclusion}.}

\section{System Model} \label{sec:SystemModel}
\begin{figure}
	\centering
	\includegraphics[scale=.52]{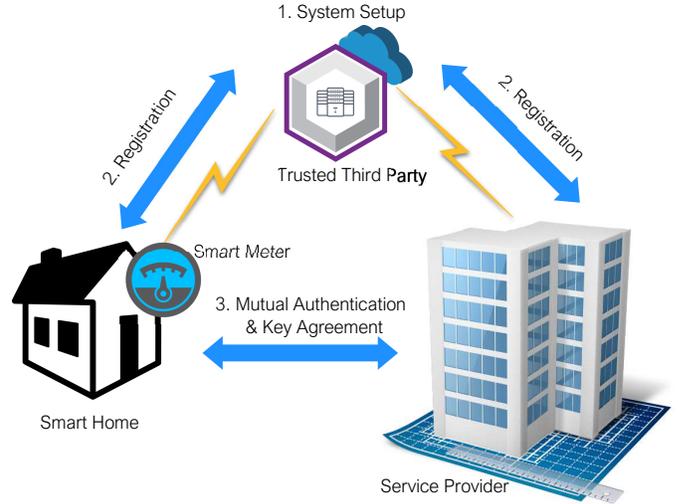}
	\caption{A typical setup of LiSA.}
	\label{fig:systemmodel}
\end{figure}

{\color{black}
A typical schematic diagram of the proposed LiSA protocol is depicted in Fig.~\ref{fig:systemmodel}. The setup comprises primarily of three entities, \ie SMs, Service Providers (SPs), and Trusted Third Party (TTP). SMs are resource constrained electronic devices deployed across customers' premises to record energy consumption on a periodic basis. Additionally, SMs  are also responsible for collecting and transmitting the energy consumption data to the utility. On the other hand, SPs are utilities that provide services to the end-users. Above all, TTP is the trusted entity that initializes the system parameters and aids in registration of SMs and SPs. Following this, mutual authentication and key exchange is executed between the SMs and SPs. The execution details of LiSA are elaborated in the upcoming segment.

}

\section{LiSA: A Mutual Authentication and Key Exchange Strategy for SMI} \label{sec:ProposedProtocol}

{\color{black} In the proposed scheme, the concepts of ECQV, ECDLP, and ECDHP are exploited to establish trust and authenticity between the SMs and SPs to resist different attack vectors. The overall execution strategy has been segregated into the following broad phases: i) Setup Phase, ii) Registration Phase, and iii) Authentication and Key Exchange Phase \cite{2019arXiv190401168K, 8718355}. In summary, the by-product of the protocol is equivalent to \textit{``mutual authentication"}, while the end-product is the \textit{``exchange of a secure session key"}. The detailed information is given below:

\subsection{Phase I: Setup Phase} During this phase, the SMI is prepared for the subsequent phases and incorporates the following course of actions:

\vspace{0.7mm}
\noindent \textit{Step 1:} It begins with the selection of an elliptic curve $E$ by the TTP. The $E$  is further associated with a generator $P$ and order $q$. 

\vspace{0.7mm}
\noindent \textit{Step 2:} Following this, the TTP computes its private ($d_{T}$) and public key ($Q_{T}$) pairs. This is done in accordance with the following equations: $d_{T} \in Z^{*}_q$ and $Q_{T}=d_{T}.P$. 

\vspace{0.7mm}
\noindent \textit{Step 3:} Next, a one-way hash function, \ie ($H_0()$) is selected.

\vspace{0.7mm}
\noindent \textit{Step 4:} Finally, the above mentioned parameters  including $<E, P, q, H_0(), Q_{T}>$ are made public.

\subsection{Phase II: Registration Phase}  During this phase, the SMs and the SPs get themselves registered with the TTP. The process of registration for both the entities is exactly alike and is based on the ECQV certificate scheme \cite{braeken2018efficient}. Thus, it is discussed in the case of SMs only, but is extendable to the SPs as well. The execution steps of the registration process are detailed in Fig.~\ref{fig:Phase2} and discussed as under:

\vspace{0.7mm}
\noindent \textit{Step 1:} The $A^{th}$ SM starts its registration process by choosing an identity ($ID_A$) for its itself. Next, it selects a random number ($r_A$) and computes $R_A$. Finally, the SM transmits $<ID_u,~R_A>$ to the TTP using the secure channel.

\vspace{0.7mm}
\noindent \textit{Step 2:} Upon receiving $<ID_u,~R_A>$, the TTP selects a random number $r_T$ and computes $R_T$. It then computes a certificate for the $A^{th}$ SM using the following operation: $Cert_A= R_A + R_T$. Following this, the TTP calculates the value of variable $r$ using $H_0()$, concatenation, and ECC point addition operations.

\vspace{0.7mm}
\noindent \textit{Step 3:} Finally, the computed value of $Cert_A$ and $r$ are sent to the $A^{th}$ SM  over the secure channel.

\vspace{0.7mm}
\noindent \textit{Step 4:} Using the received values, the SM computes its private key ($d_A$) followed by its public key $Q_A$.

\begin{figure}[t]
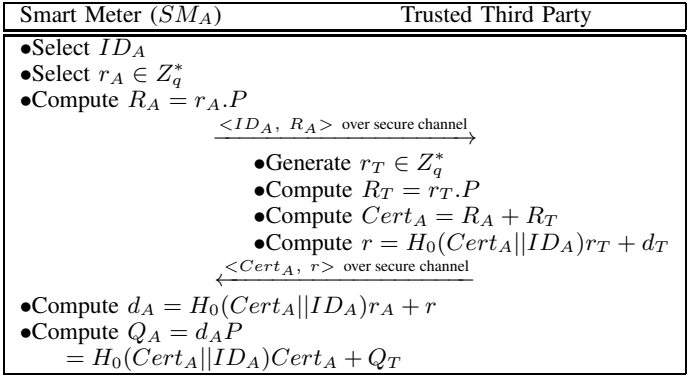

	\centering
	\small
	\begin{tabular}{|p{1.75cm} p{.5cm} p{4.4cm}|}
		\hline
		\multicolumn{1}{|c}{Smart Meter ($SM_A$)} & {}  & \multicolumn{1}{c|}{Trusted  Third Party}   \\
		\hline
		\hline
		\multicolumn{3}{|l|}{$\bullet$Select $ID_A$}\\
		\multicolumn{3}{|l|}{$\bullet$Select $r_A \in Z^{*}_q$}\\
		\multicolumn{3}{|l|}{$\bullet$Compute $R_A=r_A.P$}\\
		\multicolumn{3}{|c|}{$\xrightarrow{<ID_A,~R_A> ~\text{over secure channel}}$}\\
		&\multicolumn{2}{l|}{$\bullet$Generate $r_T \in Z^{*}_q$}\\
		&\multicolumn{2}{l|}{$\bullet$Compute $R_T=r_T.P$}\\
		&\multicolumn{2}{l|}{$\bullet$Compute $Cert_A= R_A + R_T$}\\
		&\multicolumn{2}{l|}{$\bullet$Compute $r=H_0(Cert_A||ID_A)r_T + d_{T}$}\\
		\multicolumn{3}{|c|}{ $\xleftarrow{<Cert_A,~r>~\text{over secure channel}}$}\\
		\multicolumn{3}{|l|}{$\bullet$Compute $d_A= H_0(Cert_A||ID_A)r_A + r$}\\
		\multicolumn{3}{|l|}{$\bullet$Compute $Q_A= d_AP$}\\
		\multicolumn{3}{|l|}{$\hspace*{5mm}$~$= H_0(Cert_A||ID_A)Cert_A +  Q_{T}$}\\
		\hline
	\end{tabular}
	\caption{\color{black} Phase II: Registration Phase.}
	\label{fig:Phase2}
\end{figure}

\subsection{Phase III: Authentication  \& Key Exchange Phase}  \label{sec:PhaseIII}
This is the most crucial phase of LiSA and incorporates mutual authentication and session key exchange between the SMs and SPs. The series of execution steps are reflected in Fig.~\ref{fig:Phase4} and explained as follows. 

\vspace{0.7mm}
\noindent \textit{Step 1:} The process is initiated by the $A^{th}$ SM by generating its current time stamp ($T_{SM}$) and a random number ($r_{SM}$). Next, it computes $R_{SM}$ and $R_{SM}^{'}$ using ECC point multiplication.  

\vspace{0.7mm}
\noindent \textit{Step 2:} Following this, the $A^{th}$ SM computes an intermediate token ($\mathbb{T}o_{SM}$) using its $Cert_A$ and $ID_A$. 	It also calculates an authorization token ($\mathbb{A}uth_{SM}$) for the SP to verify using the following operation: $\mathbb{A}uth_{SM}= R_{SM}^{'} \oplus \mathbb{T}o_{SM} \oplus T_{SM}$.

\vspace{0.7mm}
\noindent \textit{Step 3:} Next, $<T_{SM},~R_{SM},~\mathbb{A}uth_{SM}>$ are transmitted to $SP_B$ over the open channel. It is worth noting here that the SM's ID is never transmitted over the channel in clear text format which prevents tracking of SM by the adversary $\mathcal{A}$.

\vspace{0.7mm}
\noindent \textit{Step 4:} Once $<T_{SM},~R_{SM},~\mathbb{A}uth_{SM}>$ are received by the SP, it validates the time stamp $T_{SM}$. If found valid, the SP proceeds; else the connection is terminated immediately. Next, the SP computes $R_{SM}^{''}=d_{B}.R_{SM}$; followed by the extraction of $(Cert_{A}||ID_A)$ from the received $\mathbb{A}uth_{SM}$. Using the extracted values, the SP computes $SM_A$'s public key: $Q_A= H_0(Cert_A||ID_A)Cert_A +  Q_{T}$. If the computed key matches the publicly available $Q_A$, then the SP is sure about the authenticity of the SM. Otherwise, the connection is dropped by the SP.

\vspace{0.7mm}
\noindent \textit{Step 5:} Next, the SP records its time stamp $T_{SP}$ and generates another authentication token ($\mathbb{A}uth_{SP}$) for the SM to verify. Simultaneously, the SP also computes a session key ($SK_{B-A}$) that is used only if the mutual authentication is established. $SK_{B-A}$ is computed using the following operation: $kdf (ID_A||  Cert_A || R_{SM}^{''} ||T_{SP} || T_{SM} )$ where $kdf$ is a key derivative function.

\vspace{0.7mm}
\noindent \textit{Step 6:} Finally, the parameters $<T_{SP}, ~\mathbb{A}uth_{SP}>$ are communicated to the SM for the next course of action.

\vspace{0.7mm}
\noindent \textit{Step 7:} On receiving the above mentioned parameters, the validation of time stamp $T_{SP}$ is done.  Next, the authentication token $\mathbb{A}uth_{SP}^{*}$ is computed for verifying it against the received token. A match indicates that the SP is a valid/authentic entity. 

\vspace{0.7mm}
\noindent \textit{Step 8:} Finally, the SM computes the shared session key as follows: $SK_{A-B} = kdf (ID_A||  Cert_A || R_{SM}^{'} ||T_{SP} || T_{SM}$.

\begin{figure*}[t]
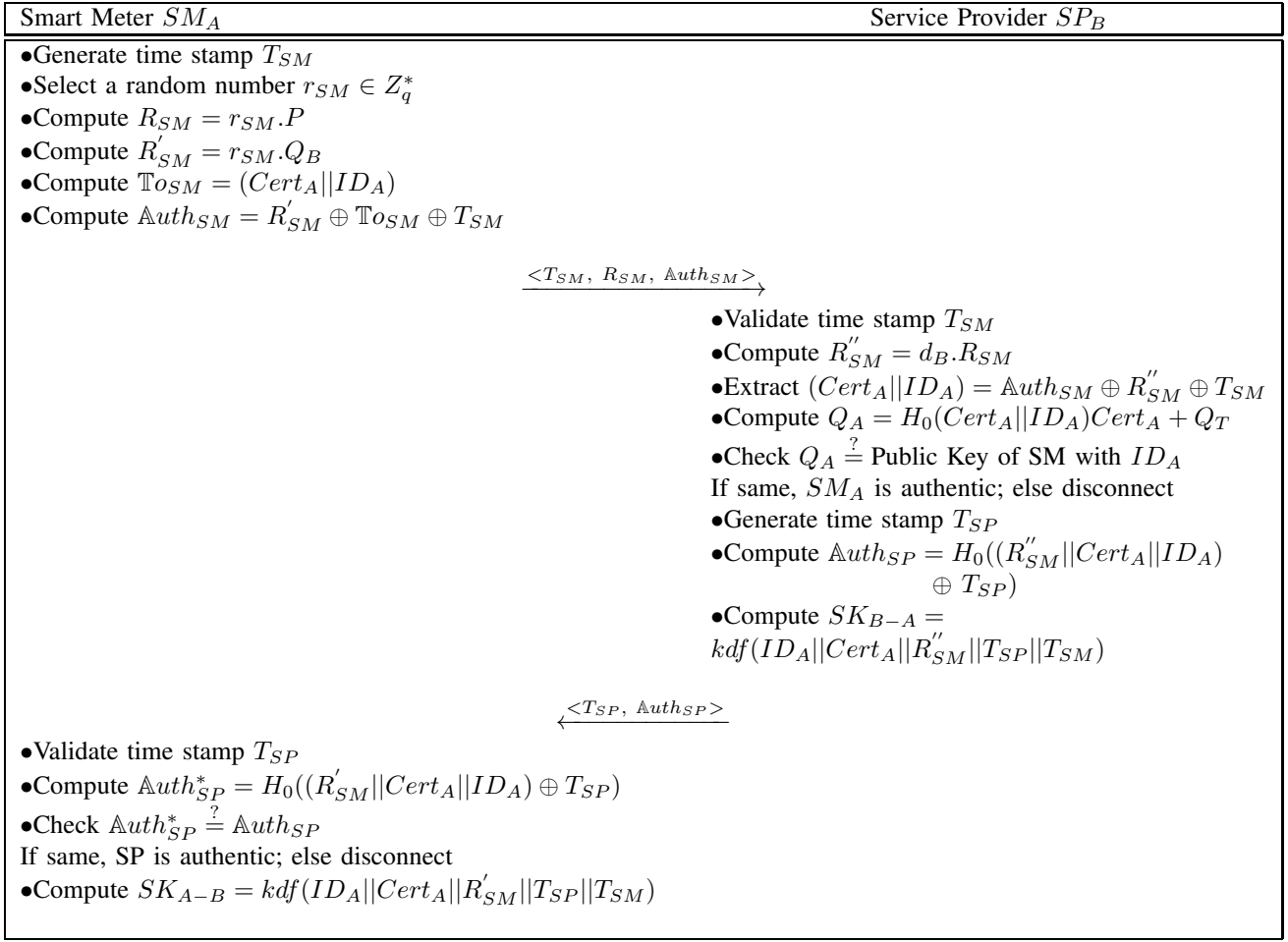

	\centering
	\begin{tabular}{|p{6cm} p{2.5cm} p{6cm}|}
		\hline
		\multicolumn{2}{|l}{{Smart Meter $SM_A$}} & \multicolumn{1}{c|}{{Service Provider $SP_B$}}   \\
		\hline
		\hline
		\multicolumn{3}{|l|}{$\bullet$Generate time stamp $T_{SM}$ }\\
		\multicolumn{3}{|l|}{$\bullet$Select a random number $r_{SM}\in Z^{*}_q$ }\\
		\multicolumn{3}{|l|}{$\bullet$Compute $R_{SM}=r_{SM}.P$ }\\
		\multicolumn{3}{|l|}{$\bullet$Compute $R_{SM}^{'}=r_{SM}.Q_{B}$ }\\
		\multicolumn{3}{|l|}{$\bullet$Compute $\mathbb{T}o_{SM}= (Cert_{A}||ID_A)$ }\\
		\multicolumn{3}{|l|}{$\bullet$Compute $\mathbb{A}uth_{SM}= R_{SM}^{'} \oplus \mathbb{T}o_{SM} \oplus T_{SM}$}\\

		&&\\
		\multicolumn{3}{|c|}{$\xrightarrow{<T_{SM},~R_{SM},~\mathbb{A}uth_{SM}>}$}\\
		
		&&\multicolumn{1}{l|}{$\bullet$Validate time stamp $T_{SM}$}\\
		&&\multicolumn{1}{l|}{$\bullet$Compute $R_{SM}^{''}=d_{B}.R_{SM}$}\\
		&&\multicolumn{1}{l|}{$\bullet$Extract $(Cert_{A}||ID_A)= \mathbb{A}uth_{SM} \oplus R_{SM}^{''} \oplus  T_{SM}$}\\
		&&\multicolumn{1}{l|}{$\bullet$Compute $Q_A= H_0(Cert_A||ID_A)Cert_A +  Q_{T} $}\\
		&&\multicolumn{1}{l|}{$\bullet$Check $Q_A \stackrel{?}{=} \text{Public Key of SM with}~ID_A$}\\
		&&\multicolumn{1}{l|}{If same, $SM_A$ is authentic; else disconnect}\\
		&&\multicolumn{1}{l|}{$\bullet$Generate time stamp $T_{SP}$}\\
		&&\multicolumn{1}{l|}{$\bullet$Compute $\mathbb{A}uth_{SP} = H_0((R_{SM}^{''} || Cert_A||ID_A)$}\\
		&&\multicolumn{1}{l|}{\hspace*{30mm}$\oplus ~T_{SP})$}\\
		&&\multicolumn{1}{l|}{$\bullet$Compute $SK_{B-A} = $}\\
		&&\multicolumn{1}{l|}{$kdf (ID_A||  Cert_A || R_{SM}^{''} ||T_{SP} || T_{SM} )$}\\
		&&\\
		\multicolumn{3}{|c|}{$\xleftarrow{<T_{SP}, ~\mathbb{A}uth_{SP}>}$}\\
		\multicolumn{3}{|l|}{$\bullet$Validate time stamp $T_{SP}$}\\
		\multicolumn{3}{|l|}{$\bullet$Compute $\mathbb{A}uth_{SP}^{*}= H_0((R_{SM}^{'} || Cert_A||ID_A) \oplus T_{SP})$}\\
		\multicolumn{3}{|l|}{$\bullet$Check $\mathbb{A}uth_{SP}^{*} \stackrel{?}{=}  \mathbb{A}uth_{SP}$ } \\
		\multicolumn{3}{|l|}{If same, SP is authentic; else disconnect}\\
		\multicolumn{3}{|l|}{$\bullet$Compute $SK_{A-B} = kdf (ID_A||  Cert_A || R_{SM}^{'} ||T_{SP} || T_{SM})$}\\
		&&\\
		\hline
		
	\end{tabular}
	\caption{\color{black}Phase III: Authentication  \& Key Exchange Phase.}
	\label{fig:Phase4}
\end{figure*}
}

\section{Security Analysis and Comparisons} \label{sec:SecurityAnalysis}

{\color{black} 
In this section, detailed security analysis of the proposed LiSA protocol is illustrated. Along with this, the comparative security assessment against the existing schemes is also presented herewith.

\subsection{Security Analysis} 
LiSA provides a number of security features and resists various attack vectors. The detailed information is illustrated below.

\subsubsection{Mutual authentication}  LiSA supports the concept of mutual authentication between the communication parties, \ie SMs and SPs. The process of mutual authentication helps to establish the authenticity of one party to the other. In the designed protocol, it is realised using two different intermediate authentication tokens namely $\mathbb{A}uth_{SM}$ and $\mathbb{A}uth_{SP}$. The hardness of these tokens can be attributed to ECQV and ECDHP \cite{braeken2018efficient}. In detail, these tokens cannot be computed without the knowledge of  $Cert_{A}, ID_A,$ and $d_B$; which are specifically private to the legitimate entities and are not shareable.

\subsubsection{Replay protection} Resistance against replay attacks enable the protocol to deny any previous message to retransmitted between the legitimate parties. In the proposed LiSA protocol, replay protection has been provided using the concept of time stamps and pseudo random numbers. These time stamps and numbers enable new values to the intermediate tokens in every session. Further, during every session, the time stamps are verified for their validity. In case of violation, the ongoing communications are terminated to avoid replay attacks. 

\subsubsection{Resistance against impersonation attacks} The proposed protocol also resists impersonation attacks. This implies that an adversary $\mathbb{A}$ cannot impersonate a valid SM to form a legitimate session with the SP. This can be credited to the strength of the ECQV certificate mechanism and the $ID$ masking mechanism that prevents an $\mathbb{A}$ to track SMs and spoof their identities. Further, extraction of a SM's certificate and identity from $\mathbb{A}uth_{SM}$ is an intractable process for an $\mathbb{A}$. This is because, these values can only be extracted by a valid SP (say $SP_B$ with which the SM is trying to connect) with its respective private key $d_B$.

\subsubsection{Resistance against man-in-the-middle (MITM) attacks} The MITM attack allows an $\mathbb{A}$ to intercept a legitimate session and spoof the identity of an entity. Following this,  $\mathbb{A}$ alters the relayed messages to illegitimately develop a session with the SM/SP. However, as explained earlier, LiSA resists impersonation and replay attacks. Further, the alteration of  $\mathbb{A}uth_{SM}$ and $\mathbb{A}uth_{SP}$ tokens is also not possible in the proposed scheme. Thus, LiSA prevents MITM attacks.

\subsubsection{Supports anonymity} LiSA also support perfect anonymity for the participating SMs. This is achieved by masking the SM's identity and certificate information while trying to develop a session with the SPs. Precisely, the SM's confidential information such as $Cert_{A}$ and $ID_A$ are never relayed in clear text format. Further, their extraction from the intermediate tokens by an illegitimate party is an intractable process. This can be attributed to the hardness of ECDLP \cite{braeken2018efficient}.

\subsubsection{Session key security} The session keys ($SK_{A-B}$ and $SK_{B-A}$) derived by the SMs and SPs in LiSA are completely secure. The session keys can only be computed by the valid and authentic parties post successful mutual authentication as discussed in Section~\ref{sec:PhaseIII}. 

\subsubsection{Resistance against  Denial of Service (DoS) attacks}
In LiSA, the process of mutual authentication is essentially initiated by the SM by relaying the following message: $\textless T_{SM},~R_{SM},~\mathbb{A}uth_{SM}\textgreater$. Post receiving this message, the SP makes checks i) the validity of the received time stamp $T_{SM}$ followed by ii) the authenticity of the SM using $\mathbb{A}uth_{SM}$. In the course of this process, if any validation check fails, then the SP knows that the other party is illegitimate. Thus, the connection is immediately dropped to prevent DoS attacks.

}

{\color{black} 
\subsection{Security Comparison}
This sections presents the detailed comparison between the proposed LiSA protocol and the existing schemes \cite{odelu2018provably, chen2017anonymous, 8294238, braeken2018efficient}. The comparison is carried out on the basis of following features: mutual authentication ($\mathfrak{SF}_1$), replay protection ($\mathfrak{SF}_2$), resistance against impersonation attacks ($\mathfrak{SF}_3$), resistance against MITM attacks  ($\mathfrak{SF}_4$),  anonymity ($\mathfrak{SF}_5$), session key security ($\mathfrak{SF}_6$), resistance against DoS attacks ($\mathfrak{SF}_7$), and lightweight ($\mathfrak{SF}_8$). 
\begin{table}[ht]
	\centering
	\caption{Analysis of security features.}
	\label{tb:ComparisionAuthentication}
	\begin{tabular}{|p{1cm} | p{1cm} |p{1cm} |p{1cm} |p{1cm} |p{1cm} |}
		\hline
		{Protocol} & \cite{odelu2018provably}  &\cite{chen2017anonymous} &  \cite{8294238}& \cite{braeken2018efficient} & {Proposed} \\
		\hline
		\hline
		$\mathfrak{SF}_1$ & $\checkmark$ &$\checkmark$&$\checkmark$&$\times$ &$\checkmark$\\
		$\mathfrak{SF}_2$ & $\checkmark$&$\checkmark$&$\checkmark$&$\checkmark$ &$\checkmark$ \\
		$\mathfrak{SF}_3$ &$\checkmark$&$\checkmark$&$\checkmark$&$\checkmark$ &$\checkmark$\\
		$\mathfrak{SF}_4$ &$\checkmark$&$\checkmark$&$\checkmark$&$\checkmark$ &$\checkmark$ \\
		$\mathfrak{SF}_5$ &$\checkmark$&$\checkmark$&$\checkmark$&$\checkmark$&$\checkmark$  \\
		$\mathfrak{SF}_6$&$\times$&$\times$&$\times$&$\checkmark$ &$\checkmark$\\
		$\mathfrak{SF}_7$ &$\times$&$\times$&$\times$&$\checkmark$ &$\checkmark$\\
		$\mathfrak{SF}_8$ &$\times$&$\times$&$\times$&$\times$ &$\checkmark$\\
		\hline
	\end{tabular}
	
\end{table}

It is evident from the comparison shown in Table~\ref{tb:ComparisionAuthentication} that the proposed LiSA protocol is the most secure out of the existing schemes. For instance, Odelu \etal's \cite{odelu2018provably}, Chen \etal's \cite{chen2017anonymous}, and Abbasinezhad-Mood \& Nikooghadam's schemes \cite{8294238} fail to provide session key security and resistance against DoS attacks. Additionally, these protocols are not lightweight in comparison to the other protocols. Finally, Braeken \etal's scheme \cite{braeken2018efficient} neither provides mutual authentication nor is lightweight.

}

\section{Performance Evaluation} \label{sec:PerformanceEvaluation}

{\color{black}
In this section, the comparative analysis of the proposed LiSA protocol is carried out against the current state-of-the-art \cite{odelu2018provably, chen2017anonymous, 8294238, braeken2018efficient}. The related comparisons have been performed on the basis of total computational and communicational complexities associated with these protocols. The detailed information about these assessments and the related simulation parameters are discussed herewith.

\subsection{Simulation Details}
Braeken \etal in \cite{braeken2018efficient} adopted an efficient means to simulate the SG setup and evaluated different protocols on it. In the said setup, the authors employed a personal computer to mimic the computational capability of the SPs; while a constrained device was used to represent a SM. The configurations of SPs included Windows 7 running on 2.5 GHz CPU and 8 GB RAM. On the other hand, SMs had limited capabilities with a single core 798 MHz CPU and 256 MB RAM.

\subsection{Evaluation Parameters} For the evaluation of the proposed LiSA protocol in comparison to the existing protocols, the following metrics have been chosen:

\subsubsection{Computational Overhead Analysis}
The computational time of the proposed and existing protocols have been computed by taking into consideration the number of computationally expensive cryptographic operations executed during the authentication and key exchange phases. Table~\ref{tbl:ComputatonalProcessingTime} illustrates the computational time  (in ms) required for executing different functions on the SMs and SPs, respectively. Here, $\mathbb{T}_{b}, \mathbb{T}_{m}, \mathbb{T}_{a}, \mathbb{T}_{h}$, $\mathbb{T}_{e}$, and $\mathbb{T}_{s}$ refer to the execution time for performing bilinear pairing, point multiplication, point addition, hashing, modular exponentiation operation, and symmetric encryption/decryption operation, respectively.

\begin{table}[h]
	\centering
	\caption{Execution time of different cryptographic functions.}
	\label{tbl:ComputatonalProcessingTime}
	\begin{tabular}{|p{.9cm}| p{.75cm}| p{.75cm}| p{.75cm}| p{.75cm}|  p{.75cm}| p{.75cm}|}
		\hline
		
		Entity & $\mathbb{T}_{b}$ & $\mathbb{T}_{m}$ & $\mathbb{T}_{a}$& $\mathbb{T}_{h}$ & $\mathbb{T}_{e}$ & $\mathbb{T}_{s}$ \\
		\hline
		\hline
		SP  & 17.01 & 0.986 & 0.004 & 0.001 & {0.874} & 0.033\\
		SM & 9.23 & 5.9 & 0.004 & 0.026 & 7.86 &0.079 \\
		\hline
	\end{tabular}
\end{table}

In accordance with the values depicted in the above table, the computational complexities of different protocols have been computed. The related outcomes are shown Table~\ref{tbl:ComputationOverhead}; wherein the overheads on the SMs and SPs are depicted separately. The proposed LiSA protocol requires 11.826 msec on the SM level and almost 0.992 msec on the SP level, for executing the authentication mechanism along with the key exchange. Additionally, the said execution time is the least in comparison with the other protocols as evidenced from the results shown in Table~\ref{tbl:ComputationOverhead}.
\begin{table}[h]
	\centering
	
	\caption{An illustration of computational complexity analysis.}
	\label{tbl:ComputationOverhead}

		\begin{tabular}{|p{2.1cm}|p{2.5cm}|p{2.5cm}|}
			\hline
			
			{Protocol} & {At SM} & {At SP }\\
			\hline
			\hline
	
			Odelu \etal \cite{odelu2018provably}  &$3\mathbb{T}_{m} + \mathbb{T}_{a}+ \mathbb{T}_{e} + 6\mathbb{T}_{h} \approx 25.72$&$2\mathbb{T}_{m} + \mathbb{T}_{a}+ 2\mathbb{T}_{b} + \mathbb{T}_{e} + 6\mathbb{T}_{h} \approx 37.28$\\
			Chen \etal \cite{chen2017anonymous} &$2\mathbb{T}_{m} + \mathbb{T}_{e} + 5 \mathbb{T}_{h} \approx 19.79$& $3\mathbb{T}_{m} + \mathbb{T}_{b} + \mathbb{T}_{e} + 5\mathbb{T}_{h} \approx 21.26$\\
			Abbasinezhad-Mood \cite{8294238} & $4\mathbb{T}_{m} + \mathbb{T}_{a} + 5 \mathbb{T}_{h} \approx 23.80$ & $4\mathbb{T}_{m} + \mathbb{T}_{a}+ 5\mathbb{T}_{h} \approx 3.98$\\
			Braeken \etal \cite{braeken2018efficient} & $4\mathbb{T}_{m}+ \mathbb{T}_{a}+\mathbb{T}_{s}+5\mathbb{T}_{h} \approx 23.81$ & $4\mathbb{T}_{m} + 2\mathbb{T}_{a}+ \mathbb{T}_{s}+ 5\mathbb{T}_{h} \approx 3.99$\\
			Proposed & $2\mathbb{T}_{m} + \mathbb{T}_{h} \approx  11.826$ & $\mathbb{T}_{m} + \mathbb{T}_{a}+ 2\mathbb{T}_{h} \approx 0.992$\\
			\hline
		\end{tabular}
\end{table}

\subsubsection{Communicational Overhead Analysis}
In order to analyse the communicational complexity of the proposed and the existing protocols, the number of messages and bits transmitted between the SMs and SPs have been considered. For this analysis, the transmitted messages have been segregated into different tokens with variable lengths. For instance, the length of the hash function and random numbers is equivalent to 160 bits, while the length of the identity field  and time stamps are set to 32 bits each. Further, the x-coordinate of an ECC point requires  60 bits. Likewise, the message length of other tokens can be found in the literature \cite{braeken2018efficient}. Consequently, the communicational overheads  associated with the transmission of different tokens across the authentication and key exchange mechanism are summarized in Table~\ref{tbl:CommunicationOverhead}. It is quite clear from the obtained results that LiSA requires the least number of rounds and number of bits to establish trust, authenticity and secure session key between the communicating parties. In detail, LiSA requires 2 rounds and 544 bits for transmission of messages between the SMs and SPs.

Thus, it can be summarized from the above discussion that LiSA is the most secure and lightweight authentication scheme for SMIs in SG setups.
}

\begin{table}[h]
	\centering
	\caption{An illustration of communicational complexity analysis.}
	\label{tbl:CommunicationOverhead}
	\begin{tabular}{|p{3cm}| p{2.2cm}| p{1.2cm}|}
		\hline
		
		Protocol & Number of Rounds & No. of bits\\
		\hline
		\hline
		Odelu \etal \cite{odelu2018provably} & 3& 1920 \\
		Chen \etal \cite{chen2017anonymous} & 3 &  1632\\
		Abbasinezhad-Mood \cite{8294238} & 3 &  832 \\
		Braeken \etal \cite{braeken2018efficient} &3 & 832\\
		Proposed & 2& 544\\
		\hline
	\end{tabular}
\end{table}

\section{Conclusion} \label{sec:Conclusion}
{\color{black}In this paper, a provably secure  mutual authentication and key agreement protocol named \textit{LiSA} was devised, specifically for SMIs employed in SG environments. The proposed protocol is a perfect mix of security features and lightweight  cryptographic functions that make it apt for deployment in SG ecosystems. In order to attain the above mentioned features, the designed protocol utilizes ECC, one-way hash functions, concatenation, and logical XOR operations. In terms of ECC, the protocol leverages the concept of ECQV, ECDLP, and ECDHP to generate the intermediate tokens including authentication tokens and session keys. Further, the detailed security assessment of the protocol reveals that it supports various security features namely mutual authentication, replay protection, anonymity, resistance against MITM, DoS, impersonation attacks, etc.  Additionally, the computational and communicational overheads associated with the execution of the designed protocol in SG setups indicate that it is relatively lightweight in comparison with the current state-of-the-art.}

\section*{Acknowledgment}
This work was supported by the Tier 2 Canada Research Chair.

\bibliographystyle{IEEEtran}
\bibliography{GCRef.bib}

\begin{thebibliography}{10}
\providecommand{\url}[1]{#1}
\csname url@samestyle\endcsname
\providecommand{\newblock}{\relax}
\providecommand{\bibinfo}[2]{#2}
\providecommand{\BIBentrySTDinterwordspacing}{\spaceskip=0pt\relax}
\providecommand{\BIBentryALTinterwordstretchfactor}{4}
\providecommand{\BIBentryALTinterwordspacing}{\spaceskip=\fontdimen2\font plus
\BIBentryALTinterwordstretchfactor\fontdimen3\font minus
  \fontdimen4\font\relax}
\providecommand{\BIBforeignlanguage}[2]{{%
\expandafter\ifx\csname l@#1\endcsname\relax
\typeout{** WARNING: IEEEtran.bst: No hyphenation pattern has been}%
\typeout{** loaded for the language `#1'. Using the pattern for}%
\typeout{** the default language instead.}%
\else
\language=\csname l@#1\endcsname
\fi
#2}}
\providecommand{\BIBdecl}{\relax}
\BIBdecl

\bibitem{kaur2018game}
K.~Kaur, S.~Garg, N.~Kumar, and A.~Y. Zomaya, ``{A Game of Incentives: An
  Efficient Demand Response Mechanism using Fleet of Electric Vehicles},'' in
  \emph{Proceedings of the 1st International Workshop on Future Industrial
  Communication Networks}.\hskip 1em plus 0.5em minus 0.4em\relax ACM, 2018,
  pp. 27--32.

\bibitem{aujla2019drops}
G.~S. Aujla, S.~Garg, S.~Batra, N.~Kumar, I.~You, and V.~Sharma, ``{DROpS: A
  Demand Response Optimization Scheme in SDN-enabled Smart Energy Ecosystem},''
  \emph{Information Sciences}, vol. 476, pp. 453--473, 2019.

\bibitem{8642293}
P.~{Kumar}, Y.~{Lin}, G.~{Bai}, A.~{Paverd}, J.~S. {Dong}, and A.~{Martin},
  ``{Smart Grid Metering Networks: A Survey on Security, Privacy and Open
  Research Issues},'' \emph{IEEE Communications Surveys \& Tutorials}, 2019,
  {DOI: 10.1109/COMST.2019.2899354}.

\bibitem{7959167}
M.~R. {Asghar}, G.~{Dán}, D.~{Miorandi}, and I.~{Chlamtac}, ``{Smart Meter
  Data Privacy: A Survey},'' \emph{IEEE Communications Surveys \& Tutorials},
  vol.~19, no.~4, pp. 2820--2835, 2017.

\bibitem{kaursystem}
K.~Kaur, S.~Garg, N.~Kumar, G.~S. Aujla, K.~K.~R. Choo, and M.~S. Obaidat,
  ``{An Adaptive Grid Frequency Support Mechanism for Energy Management in
  Cloud Data Centers},'' \emph{IEEE Systems Journal}, 2019, {DOI:
  10.1109/JSYST.2019.2921592}.

\bibitem{kaur2019osco}
K.~Kaur, S.~Garg, G.~Kaddoum, S.~H. Ahmed, and D.~N.~K. Jayakody, ``{En-OsCo:
  Energy-aware Osmotic Computing Framework using Hyper-heuristics},'' in
  \emph{Proceedings of the ACM MobiHoc Workshop on Pervasive Systems in the IoT
  Era}.\hskip 1em plus 0.5em minus 0.4em\relax ACM, 2019, pp. 19--24.

\bibitem{chaudhary2018sdn}
R.~Chaudhary, G.~S. Aujla, S.~Garg, N.~Kumar, and J.~J. Rodrigues,
  ``{SDN-enabled Multi-Attribute-based Secure Communication for Smart Grid in
  IIoT Environment},'' \emph{IEEE Transactions on Industrial Informatics},
  vol.~14, no.~6, pp. 2629--2640, 2018.

\bibitem{7093120}
S.~{Finster} and I.~{Baumgart}, ``{Privacy-Aware Smart Metering: A Survey},''
  \emph{IEEE Communications Surveys \& Tutorials}, vol.~17, no.~2, pp.
  1088--1101, 2015.

\bibitem{2019arXiv190401171G}
S.~{Garg}, K.~{Kaur}, G.~{Kaddoum}, F.~{Gagnon}, and J.~J. P.~C. {Rodrigues},
  ``{An Efficient Blockchain-Based Hierarchical Authentication Mechanism for
  Energy Trading in V2G Environment},'' in \emph{IEEE International Conference
  on Communications Workshops (ICC Workshops), Shanghai, China}, May 2019.

\bibitem{8676005}
A.~{Ghosal} and M.~{Conti}, ``{Key Management Systems for Smart Grid Advanced
  Metering Infrastructure: A Survey},'' \emph{IEEE Communications Surveys \&
  Tutorials}, pp. 1--1, 2019, {DOI: 10.1109/COMST.2019.2907650}.

\bibitem{diao2015privacy}
F.~Diao, F.~Zhang, and X.~Cheng, ``{A Privacy-Preserving Smart Metering Scheme
  using Linkable Anonymous Credential},'' \emph{IEEE Transactions on Smart
  Grid}, vol.~6, no.~1, pp. 461--467, 2015.

\bibitem{7138617}
K.~{Yu}, M.~{Arifuzzaman}, Z.~{Wen}, D.~{Zhang}, and T.~{Sato}, ``{A Key
  Management Scheme for Secure Communications of Information Centric Advanced
  Metering Infrastructure in Smart Grid},'' \emph{IEEE Transactions on
  Instrumentation and Measurement}, vol.~64, no.~8, pp. 2072--2085, 2015.

\bibitem{7931655}
D.~{Abbasinezhad-Mood} and M.~{Nikooghadam}, ``{An Ultra-Lightweight and Secure
  Scheme for Communications of Smart Meters and Neighborhood Gateways by
  Utilization of an ARM Cortex-M Microcontroller},'' \emph{IEEE Transactions on
  Smart Grid}, vol.~9, no.~6, pp. 6194--6205, 2018.

\bibitem{seferian2018identity}
V.~Seferian, R.~Kanj, A.~Chehab, and A.~Kayssi, ``{Identity based Key
  Distribution Framework for Link Layer Security of AMI Networks},'' \emph{IEEE
  Transactions on Smart Grid}, vol.~9, no.~4, pp. 3166--3179, 2018.

\bibitem{8662558}
Y.~{Lee}, E.~{Hwang}, and J.~{Choi}, ``{A Unified Approach for Compression and
  Authentication of Smart Meter Reading in AMI},'' \emph{IEEE Access}, vol.~7,
  pp. 34\,383--34\,394, 2019.

\bibitem{mustapa2018hardware}
M.~Mustapa, M.~Y. Niamat, A.~P.~D. Nath, and M.~Alam, ``{Hardware-Oriented
  Authentication for Advanced Metering Infrastructure},'' \emph{{IEEE
  Transactions on Smart Grid}}, vol.~9, no.~2, pp. 1261--1270, 2018.

\bibitem{7676395}
A.~{Mohammadali}, M.~{Sayad Haghighi}, M.~H. {Tadayon}, and
  A.~{Mohammadi-Nodooshan}, ``{A Novel Identity-Based Key Establishment Method
  for Advanced Metering Infrastructure in Smart Grid},'' \emph{IEEE
  Transactions on Smart Grid}, vol.~9, no.~4, pp. 2834--2842, 2018.

\bibitem{8413131}
P.~{Kumar}, A.~{Gurtov}, M.~{Sain}, A.~{Martin}, and P.~H. {Ha}, ``{Lightweight
  Authentication and Key Agreement for Smart Metering in Smart Energy
  Networks},'' \emph{IEEE Transactions on Smart Grid}, vol.~10, no.~4, pp.
  4349--4359, 2019.

\bibitem{8294238}
D.~{Abbasinezhad-Mood} and M.~{Nikooghadam}, ``{An Anonymous ECC-Based
  Self-Certified Key Distribution Scheme for the Smart Grid},'' \emph{IEEE
  Transactions on Industrial Electronics}, vol.~65, no.~10, pp. 7996--8004,
  2018.

\bibitem{braeken2018efficient}
A.~Braeken, P.~Kumar, and A.~Martin, ``{Efficient and Provably Secure Key
  Agreement for Modern Smart Metering Communications},'' \emph{Energies},
  vol.~11, no.~10, p. 2662, 2018.

\bibitem{2019arXiv190401168K}
K.~{Kaur}, S.~{Garg}, G.~{Kaddoum}, F.~{Gagnon}, and S.~H. {Ahmed},
  ``{Blockchain-Based Lightweight Authentication Mechanism for Vehicular Fog
  Infrastructure},'' in \emph{IEEE International Conference on Communications
  Workshops (ICC Workshops), Shanghai, China}, May 2019.

\bibitem{8718355}
S.~{Garg}, K.~{Kaur}, G.~{Kaddoum}, S.~H. {Ahmed}, and D.~N.~K. {Jayakody},
  ``{SDN based Secure and Privacy-preserving Scheme for Vehicular Networks: A
  5G Perspective},'' \emph{IEEE Transactions on Vehicular Technology}, 2019,
  {DOI: 10.1109/TVT.2019.2917776}.

\bibitem{odelu2018provably}
V.~Odelu, A.~K. Das, M.~Wazid, and M.~Conti, ``{Provably Secure Authenticated
  Key Agreement Scheme for Smart Grid},'' \emph{IEEE Transactions on Smart
  Grid}, vol.~9, no.~3, pp. 1900--1910, 2018.

\bibitem{chen2017anonymous}
Y.~Chen, J.-F. Mart{\'\i}nez, P.~Castillejo, and L.~L{\'o}pez, ``{An Anonymous
  Authentication and Key Establish Scheme for Smart Grid: FAuth},''
  \emph{Energies}, vol.~10, no.~9, p. 1354, 2017.

\end{thebibliography}

\end{document}